\documentclass[letterpaper,aps,pre,twocolumn,showcaps,superscriptaddress,showpacs,amsmath,amssymb]{revtex4}
\usepackage{graphicx}
\usepackage{color}

\newcommand{\dd}{\mathrm{d}}

\newcommand{\fd}[2]{\frac{\delta #1}{\delta #2}}
\newcommand{\mean}[1]{\langle #1 \rangle}
\newcommand{\Int}[1]{\int\dd #1\;}
\newcommand{\Intns}[1]{\int\dd #1}
\newcommand{\IInt}[3]{\int_{#2}^{#3}\dd #1\;}

\newcommand{\FInt}[1]{\int[\dd #1]\;}
\newcommand{\FIntns}[1]{\int[\dd #1]}

\renewcommand{\vec}[1]{\mathbf #1}

\newcommand{\im}{\text i}
\newcommand{\id}{\mathbf 1}
\newcommand{\x}{\vec r}         

\newcommand{\vu}{\vec u}
\newcommand{\vv}{\vec v}
\newcommand{\vw}{\boldsymbol\omega}
\newcommand{\cu}{\boldsymbol\phi}
\newcommand{\mGam}{\boldsymbol\Gamma}
\newcommand{\A}{\mathcal S}
\newcommand{\mchi}{\boldsymbol\chi}

\newcommand{\mchis}{\boldsymbol\chi_\text{s}}
\newcommand{\vsi}{\boldsymbol\psi}

\newcommand{\om}{\omega}

\begin{document}


\title{Gaussian field theory for the Brownian motion of a solvated particle}

\author{Thomas Speck}
\affiliation{Institut f\"ur Theoretische Physik II,
  Heinrich-Heine-Universit\"at D\"usseldorf, Universit\"atsstra\ss e 1, 40225
  D\"usseldorf, Germany}

\begin{abstract}
  An alternative derivation of Brownian motion is presented. Instead of
  supplementing the linearized Navier-Stokes equation with a fluctuating
  force, we directly assume a Gaussian action functional for solvent velocity
  fluctuations. Solvating a particle amounts to expelling the solvent and
  prescribing a boundary condition to the solvent on the interface that is
  shared with the solute. We study the dynamical effects of this boundary
  condition on the solvent and derive explicit expressions for the solvent
  mean flow and velocity correlations. Moreover, we show that the probability
  to observe solvent velocity fluctuations that are compatible with the
  boundary condition reproduces random Brownian motion of the solvated
  particle. We explicitly calculate the translational and rotational diffusion
  coefficients of a spherical particle using the presented formalism.
\end{abstract}


\maketitle


\section{Introduction}

Our traditional understanding of Brownian motion is that of a solute getting
kicked randomly by surrounding solvent molecules. Einstein's seminal
contribution had been to derive the diffusion equation--and in particular an
explicit expression for the diffusion coefficient--based on this
insight~\cite{eins05}. One route to derive Brownian motion from more
fundamental equations is fluctuating hydrodynamics~\cite{landau6}, which
augments the linearized Navier-Stokes equations with a random stress tensor
obeying the fluctuation-dissipation theorem. Integrating out the fluid
velocity field, the stochastic equation of motion for a solvated particle can
then be derived, which takes on the form of a generalized Langevin
equation~\cite{haug73}. This approach is not without problems since the
correlations of the random stress tensor near the solute are assumed to be the
same as those in the isotropic bulk fluid. There is, however, evidence that
the presence of rigid bodies influences the local stress
fluctuations~\cite{schi10}.

The purpose of this Brief Report is to give a somewhat alternative view on the
dynamics of solvated particles, a view in which Brownian motion arises from
suppressing solvent fluctuations. Instead of the stress, we consider directly
the (memoryless) fluctuations of the solvent velocity. We employ Gaussian
field theory~\cite{chan93} to study how a solute modifies these velocity
fluctuations in its vicinity. Gaussian field theory has been applied
successfully to, e.g., dielectric relaxation dynamics~\cite{song96} and the
hydrophobic effect~\cite{lum99}. In analogy to the modification of density
fluctuations due to the excluded volume of the solute, we show that imposing a
solvent velocity on the solute-solvent interface leads to a flow and
diminishes solvent velocity fluctuations in the vicinity of the
solute. Loosely speaking, the solvent loses entropy, which is compensated by
the random motion of the solute.


\section{The pure solvent}

We consider an incompressible quiescent fluid with bulk viscosity $\eta$. We
assume that in the absence of any solvated object the probability
$P_0[\vu]=\exp\{-\A_0[\vu]\}$ to observe a given history of velocity
fluctuations $\vu(\x,t)$ away from zero is Gaussian with action
\begin{equation}
  \label{eq:S0}
  \A_0[\vu] = \frac{1}{2} \Intns{t}\Int{^3\x\dd^3\x'}
  \vu(\x,t)\cdot\mchi_0^{-1}(\x,\x')\cdot\vu(\x',t).
\end{equation}
The functional inverse of $\mchi_0$ is defined through
\begin{equation}
  \label{eq:inv}
  \Int{^3\x''} \mchi_0^{-1}(\x,\x'')\mchi_0(\x'',\x') = \id\delta(\x-\x'),
\end{equation}
where $\id$ denotes the identity matrix. Normalization of $P_0$ is achieved
through choosing the appropiate functional measure $[\dd\vu]$. Clearly, the
matrix $\mchi_0$ is related to the velocity correlations,
\begin{equation}
  \label{eq:corr}
  \mean{\vu(\x,t)\vu^T(\x',t')}_0 = \mchi_0(\x,\x')\delta(t-t').
\end{equation}
The brackets denote averages over different realizations of the velocity
field, whereas the subscript indicates the pure solvent. Eq.~(\ref{eq:corr})
implies temporally uncorrelated velocity fluctuations. We know that this is
not strictly correct as hydrodynamics predicts a power law tail for the
particle velocity autocorrelation function~\cite{wido71,fran11}. However, for
the sake of simplicity, here we aim for a simple Markovian description which
is appropriate for sufficiently coarse-grained time. Brownian motion in
confined geometries such as nanopores~\cite{detc12} might require to take into
account memory.


Now image that an external force density $\vec f(\x,t)$ acts on the solvent.
Up to linear order, the fluctuation-dissipation theorem~\cite{kubo} relates
this force to an instantaneous mean velocity profile
\begin{equation}
  \label{eq:lr}
  \mean{\vu(\x,t)} = \frac{1}{2T}\Int{^3\x'}\mchi_0(\x,\x')\cdot\vec f(\x',t)
\end{equation}
through the correlations Eq.~\eqref{eq:corr}. The factor $1/2$ originates from
the time integration over the $\delta$-function. Of course, Eq.~\eqref{eq:lr}
is nothing else than Faxen's theorem, from which we can deduce the
correlations
\begin{equation}
  \label{eq:oseen}
  \mchi_0(\x) = \frac{T}{4\pi\eta r}\left(\id+\frac{\x\x^T}{r^2}\right)
\end{equation}
using the Oseen tensor~\cite{dhont}. However, we will not need an explicit
expression for $\mchi_0$ in the following.


\section{Solvating a particle}

We solvate a particle expelling the fluid from the volume $V$ the particle
occupies. For the clarity of presentation, we consider a spherical
particle. Then rotational and translational diffusion decouple and in the
following we focus on the translational diffusion. The particle is described
by its instantaneous velocity $\vv$. By choosing the velocity correlations
Eq.~\eqref{eq:corr} we assume a time scale separation between the relaxation
time of the fluid and the time scale on which the solvated particle
diffuses. From the perspective of the solute it thus appears as if the solvent
responds \textit{instantaneously} with a mean flow profile
$\mean{\vu(\x)}_\vv$ to a change of $\vv$.

To calculate this flow profile, we exploit the time scale separation through
fixing $\vv$ and consider the generating function
\begin{equation}
  \label{eq:Q:def}
  Q[\cu] \equiv \FIntns{\vu}\FInt{\vsi} e^{-\A_0+\im
    C+\Intns{t}\Int{^3\x}\cu(\x,t)\cdot\vu(\x,t)}
\end{equation}
with conjugate field $\cu(\x,t)$. We implement the ``no-slip'' or ``stick''
boundary condition such that the fluid layer on the particle surface has the
same velocity as the moving particle. The boundary condition gives rise to the
constraint
\begin{equation*}
  \prod_t \delta(\vu(\x,t)-\vec v(t)) = \FInt{\vsi} e^{\im C[\vsi]}
\end{equation*}
in Eq.~\eqref{eq:Q:def} with functional
\begin{equation}
  \label{eq:C}
  C[\vsi,\vu|\vec v] \equiv \Intns{t}\Int{^3\x}
  [\vu(\x,t)-\vec v(t)]\cdot\vsi(\x,t).
\end{equation}
The auxiliary function $\vsi(\x,t)$ is non-zero only on the particle surface
$\delta V$ and zero elsewhere.

We evaluate the path integrals in Eq.~\eqref{eq:Q:def} in two steps. First,
integration over the velocity fluctuations results in
\begin{widetext}
  \begin{equation}
    \label{eq:Q:1}
    Q[\cu] = \FInt{\vsi} \exp\left\{
      -\frac{1}{2}\Int{t}\vsi\circ\mchis\circ\vsi +
      \im\Int{t}\vsi\circ\vec b + \frac{1}{2}\Intns{t}\Int{^3\x\dd^3\x'}
      \cu(\x,t)\cdot\mchi_0(\x,\x')\cdot\cu(\x',t)
    \right\}.
  \end{equation}
\end{widetext}
To ease the notational burden, we define the operation
\begin{equation}
  \label{eq:op}
  \vec g\circ\vec h \equiv \IInt{^2\x}{\delta V}{} \vec g^T(\x)\vec h(\x)
\end{equation}
as the integral over all positions that are elements of the particle
surface. The matrix $\mchis$ quantifies fluid velocity correlations in the
absence of, but on the interface the fluid would share with, the solvated
particle. The linear term in Eq.~\eqref{eq:Q:1} couples to the vector
\begin{equation}
  \label{eq:b}
  \vec b(\x,t) \equiv \Int{^3\x'} \mchi_0(\x,\x')\cdot\cu(\x',t) 
  - \vec v(t),
\end{equation}
where $\x\in\delta V$. Performing the integration over the auxiliary vector
field $\vsi$ we finally obtain the generating function
\begin{multline}
  \label{eq:Q}
  Q[\cu] = \mathcal N
  \exp\left\{-\frac{1}{2}\Int{t}\vec b\circ\mchis^{-1}\circ\vec b \right. \\
  \left.
    + \frac{1}{2}\Intns{t}\Int{^3\x\dd^3\x'}
    \cu(\x,t)\cdot\mchi_0(\x,\x')\cdot\cu(\x',t)
  \right\}
\end{multline}
with (irrelevant) prefactor $\mathcal N$ due to the functional determinant of
$\mchis$. The inverse matrix is determined through
\begin{equation}
  \label{eq:invs}
  \mchis(\x,\x'')\circ\mchis^{-1}(\x'',\x') = \id\delta(\x-\x'), \quad
  \x,\x'\in\delta V,
\end{equation}
i.e., the inversion is carried out on the subspace defined by the particle
surface.


\begin{figure}[t]
  \centering
  \includegraphics[width=\linewidth]{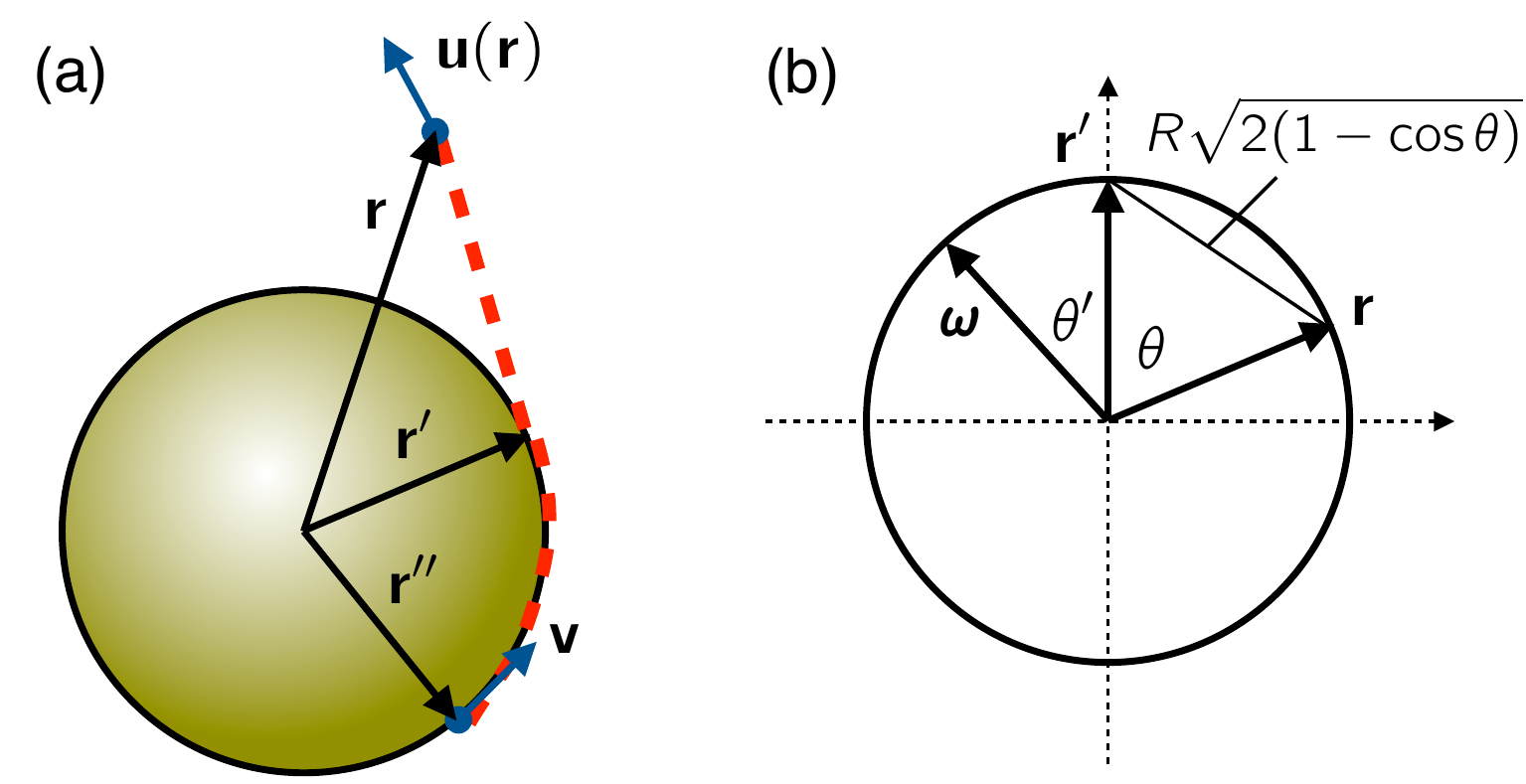}
  \caption{(a)~Schematic representation of Eq.~\eqref{eq:flow}: The flow $\vu$
    around the spherical particle at $\x$ is connected to the flow on the
    surface (given by the particle velocity $\vv$) through a concatenation of
    the tensors $\mchis^{-1}$ and $\mchi_0$. (b)~Coordinate system and symbols
    used in the appendix: $\x$ and $\x'$ are two points on the particle
    surface and $\vw$ is the angular velocity.}
  \label{fig:solv}
\end{figure}

From the generating function Eq.~\eqref{eq:Q} it is straightforward to
calculate the mean flow profile as the functional derivative
\begin{equation}
  \mean{u_i(\x,t)}_\vv = \left.\fd{\ln Q}{\phi_i(\x,t)}\right|_{\cu=0}
\end{equation}
with respect to the conjugate field. Using
\begin{equation}
  \fd{b_i(\x',t')}{\phi_j(\x,t)} = [\mchi_0(\x',\x)]_{ij}\delta(t-t')
\end{equation}
together with the symmetry property $\mchi_0(\x',\x)=\mchi_0(\x,\x')$ we
obtain
\begin{equation}
  \label{eq:flow}
  \mean{\vu(\x)}_\vv = \mchi_0(\x,\x')\circ\mchis^{-1}(\x',\x'')\circ\vv.
\end{equation}
For $\x\in\delta V$, we can use Eq.~\eqref{eq:invs} to show that
$\mean{\vu(\x)}_\vv=\vv$ as required. Fig.~\ref{fig:solv}(a) shows
schematically how the imposed surface velocity generates the flow at $\x$ in
the solvent through a concatenation of the tensors $\mchis^{-1}$ and
$\mchi_0$.

The presence of the solute not only causes a flow but also alters the solvent
velocity fluctuations. The correlations are calculated as
\begin{equation}
  [\mchi(\x,\x')]_{ij}
  = \left.\fd{^2\ln Q}{\phi_i(\x,t)\delta\phi_j(\x',t)}\right|_{\cu=0}
\end{equation}
and read
\begin{multline}
  \mchi(\x,\x') = \mean{\delta\vu(\x)\delta\vu^T(\x')}_\vv = \\
  \mchi_0(\x,\x') -
  \mchi_0(\x,\x'')\circ\mchis^{-1}(\x'',\x''')\circ\mchi_0(\x''',\x') 
\end{multline}
for $\delta\vu(\x)\equiv\vu(\x)-\mean{\vu(\x)}_\vv$. For either $\x\in\delta
V$ or $\x'\in\delta V$, i.e., directly on the particle surface, the
correlations vanish. Far away from the particle in the bulk they become those
of the pure solvent, $\mchi=\mchi_0$.


\section{Brownian motion}

Comparing the result Eq.~\eqref{eq:flow} with the linear response relation
Eq.~\eqref{eq:lr}, we see that it corresponds to a force density that is
confined to the solute-solvent interface and given by
\begin{equation}
  \label{eq:f}
  \vec f(\x) = 2T\mchis^{-1}(\x,\x')\circ\vec v, \qquad \x\in\delta V.
\end{equation}
The total force exerted by the particle follows from integrating this force
density over the particle surface,
\begin{equation}
  \label{eq:Gam}
  \vec F = \id\circ\vec f = 2T(\id\circ\mchis^{-1}\circ\id)\cdot\vec v
  \equiv \mGam\cdot\vec v.
\end{equation}
This is a linear relation between force and velocity, allowing us to identify
$\mGam$ as the friction tensor. In appendix~\ref{sec:trans}, $\mGam$ is
calculated explicitly from Eq.~\eqref{eq:Gam} for a spherical particle.

So far, we have treated the particle velocity $\vv$ as given. Inspecting
Eq.~\eqref{eq:Q:def}, we see that for $\cu=0$ the generating function reduces
to the path probability to observe spontaneous solvent velocity fluctuations
that are compatible with the boundary condition of a uniform velocity on the
solute surface. Hence, $P[\vec v(t)]\sim Q[\cu=0]$ is the path weight of a
history of particle velocities with
\begin{equation}
  \label{eq:A}
  P[\vec v(t)] \equiv
  \exp\left\{-\frac{1}{4}\Int{t}\vec v(t)\cdot\vec D^{-1}\cdot\vec v(t)
  \right\},
\end{equation}
which corresponds to the path weight of a free, overdamped Brownian
particle. The diffusion tensor
\begin{equation*}
  \vec D^{-1} \equiv 2(\id\circ\mchis^{-1}\circ\id) = \mGam/T
\end{equation*}
is given by the well-known Einstein relation connecting diffusion to friction
through the fluid temperature.

\section{Conclusions}

The formalism introduced here is not restricted to spherical particles but can
be used to obtain the friction tensors for arbitrarily shaped objects, e.g.,
colloidal clusters with complex shapes~\cite{kraf13}. Its practical use is
somewhat hampered by the difficulties to determine $\mchis^{-1}$ explicitly,
which amounts to the inversion of $\mchi_0$ on the two dimensional
sub-manifold of the solute surface. Except for simple shapes
(cf. appendix~\ref{sec:trans}) this has to be done numerically, possibly
through an expansion of $\mchis$ into a suitable matrix basis. Rotational
diffusion, and its coupling to translational diffusion, can be treated
straightforwardly as demonstrated in appendix~\ref{sec:rot}. Another advantage
of the presented formalism is that we are not restricted to use the
correlations Eq.~\eqref{eq:oseen} holding for an unbounded solvent. For
example, one might calculate the \emph{pure} solvent correlations
$\mchi_0(\x,\x')$ in some geometry numerically (e.g., using fluctuating
lattice Boltzmann simulations~\cite{adhi05}) and use these correlations to
obtain the diffusion coefficients of differently shaped objects moving in this
geometry.


\acknowledgments

I thank David Chandler for introducing me to Gaussian field theory. I
acknowledge financial support from the Alexander-von-Humboldt foundation.


\appendix

\section{Friction tensor of a spherical particle}
\label{sec:trans}

To demonstrate the validity of the present formalism, we explicitly calculate
the friction tensor for a spherical particle with radius $R$ via
Eq.~\eqref{eq:Gam}. We start by calculating the matrix
\begin{equation*}
  \vec X \equiv \id\circ\mchis = \IInt{^2\x}{\delta V}{} \mchi_0(\x-\x'),
\end{equation*}
which is the integral of the free correlations Eq.~\eqref{eq:oseen} over the
surface of the sphere with $\x'$ hold fixed. Without loss of generality, we
place the particle center at the origin and use spherical coordinates with
$\x'$ pointing along the $z$-axis, see Fig.~\ref{fig:solv}(b). The
off-diagonal components vanish after integration over the azimuth
angle. Substituting $z=\cos\theta$, the diagonal components read
\begin{gather*}
  X_{xx} = X_{yy} = \frac{TR}{2\eta} \IInt{z}{-1}{1}
  \frac{1}{\sqrt{2(1-z)}}\left[ 1+\frac{1-z^2}{4(1-z)} \right], \\
  X_{zz} = \frac{TR}{2\eta} \IInt{z}{-1}{1}
  \frac{1+\frac{1}{2}(1-z)}{\sqrt{2(1-z)}},
\end{gather*}
and therefore
\begin{equation*}
  \vec X = \frac{4TR}{3\eta}\id.
\end{equation*}
As expected, this matrix does not depend on the vector $\x'$ due to the
spherical symmetry. Following Eq.~\eqref{eq:invs}, we find
$\mchis^{-1}\circ\id=\vec X^{-1}$ and thus recover Stokes' expression
\begin{equation*}
  \mGam = 2T(\id\circ\mchis^{-1}\circ\id) = 2T\; 4\pi R^2\; \vec X^{-1} =
  6\pi\eta R\id
\end{equation*}
for the friction tensor of a sphere.

\section{Rotational diffusion}
\label{sec:rot}

In order to include rotational diffusion, we have to modify the boundary
condition. The fluid velocity on the surface now reads
\begin{equation*}
  \vu(\x,t) = \vv(t) + \vw(t) \times \x, \qquad \x\in\delta V,
\end{equation*}
where $\vw=\om\vec e$ is the angular velocity with speed $\om$ about an axis
given by the normalized vector $\vec e$. Repeating the steps leading to
Eq.~\eqref{eq:A}, we obtain the path weight $P\sim e^{-A}$ with stochastic
action
\begin{equation*}
  A = \frac{1}{2}\Int{t}
  [\vv+\vw\times\x]\circ\mchis^{-1}(\x,\x')\circ[\vv+\vw\times\x'].
\end{equation*}

We now calculate the rotational diffusion coefficient for a spherical particle
with radius $R$. We use a strategy similar to the previous section by
multiplying Eq.~\eqref{eq:invs} by the vector $\vec e\times\x$ from the left,
and $\vec e\times\x'$ from the right side followed by integrations,
\begin{equation*}
  \IInt{^2\x}{\delta V}{} \vec L(\x)\cdot\vec L^{-1}(\x)
  = \IInt{^2\x}{\delta V}{} (\vec e\times\x)\cdot(\vec e\times\x)
  = \frac{8\pi}{3}R^4.
\end{equation*}
Here, we have defined the vector field
\begin{equation*}
  \vec L(\x') \equiv
  \IInt{^2\x}{\delta V}{} \mchi_0(\x'-\x)\cdot(\vec e\times\x) =
  \frac{2}{3}\frac{TR}{\eta}\vec e\times\x'.
\end{equation*}
It is convenient to calculate the integral using spherical coordinates as
sketched in Fig.~\ref{fig:solv}(b), where the $z$-axis points along $\x'$. For
the inverse field we thus find
\begin{equation*}
  \vec L^{-1}(\x') \equiv \IInt{^2\x}{\delta V}{}
  \mchis^{-1}(\x',\x)\cdot(\vec e\times\x) = \frac{\eta}{TR}\vec e\times\x'.
\end{equation*}
Using this vector, the stochastic action can be written
\begin{equation*}
  A = \Int{t} \left\{ \frac{1}{4}\vec v\cdot\vec D^{-1}\cdot\vec v + 
  \vec v\cdot(\id\circ\vec L^{-1})\om + \frac{\Gamma_r}{4T}\om^2 \right\}.
\end{equation*}
Clearly, $\id\circ\vec L^{-1}=0$, i.e., translational and rotational diffusion
decouple. The rotational diffusion coefficient reads
\begin{equation*}
  \Gamma_r = 2T\IInt{^2\x}{\delta V}{} (\vec e\times\x)\cdot\vec L^{-1}(\x)
  = 8\pi\eta R^3
\end{equation*}
as expected.


\end{document}